\documentclass[12pt]{article}
\usepackage{fullpage}
\usepackage{epsfig}
\usepackage{amsmath}
\usepackage{amssymb}
\usepackage{color}

\begin{document}

\begin{flushright}
\end{flushright}

\centerline{\bf\Large Effects of staggered fermions and mixed actions}

\vspace{0.2cm}

\centerline{\bf\Large on the scalar correlator}

\vspace{1cm}

\centerline{\large S. Prelovsek}

\vspace{0.5cm}

\centerline{\small e-mail: {\it sasa.prelovsek@ijs.si}}

\vspace{0.5cm} 

\centerline{\small \it Department of Physics, University of Ljubljana, Jadranska 19,  1000 Ljubljana, Slovenia}

\centerline{\small \it and}

\centerline{\small \it Institute Jozef Stefan, Jamova 39,  1000 Ljubljana, Slovenia}

\vspace{1cm} 

\centerline{\bf Abstract}

\vspace{0.3cm}

We provide the analytic predictions for the flavor non-singlet scalar correlator, which will enable determination of the scalar meson mass from the lattice scalar correlator. We consider  
 simulations with 2+1 staggered sea quarks and staggered or 
chiral valence quarks.  At small $u/d$ masses the  
correlator is dominated by the bubble contribution, which 
is the intermediate state with two pseudoscalar mesons.  We determine the 
bubble contribution within
Staggered  and Mixed Chiral Perturbation Theory.
 Its effective mass is smaller than the mass of $\pi\eta$, which is the lightest intermediate state in proper 2+1 QCD. 
The unphysical effective mass is a consequence of the taste breaking that
makes possible the intermediate state with mass $2M_\pi$.  We find 
that the scalar correlator can be negative in the simulations 
with mixed quark actions if the sea and valence quark masses are tuned 
by matching the pion masses $M_{val,val}=M_{\pi_5}$.

\section{Introduction}

The Nature of the lightest observed scalar resonances is still 
not revealed yet. In this paper we consider the flavor non-singlet 
scalar meson and the problems related to its simulation on the lattice. 
Below $2$ GeV there are two experimentally well established scalar resonances 
with isospin $I=1$: $a_0(980)$ and $a_0(1450)$ \cite{pdg}. 
It is 
still not clear which of the two corresponds to the lightest $\bar qq$ 
scalar meson. 
Most of the models \cite{models} and lattice simulations 
have difficulty relating 
the mass and decay properties of $a_0(980)$ with the 
$\bar qq$ state. If $a_0(980)$ belongs to the lightest $\bar qq$ 
nonet, it does not have an obvious strange partner since the 
mass of $K_0(1430)$ 
is relatively high, while $\kappa(800)$ is experimentally very controversial 
\cite{pdg,bes} and probably too light to be the partner. This suggests the 
possibility that $a_0(1450)$ 
might be the lightest $\bar qq$ state with $I=1$, while $a_0(980)$ could be 
exotic state such as
 tetraquark $\bar q\bar qqq$ \cite{jaffe} or mesonic molecule.

\begin{figure}[htb!]
\begin{center}
\epsfig{file=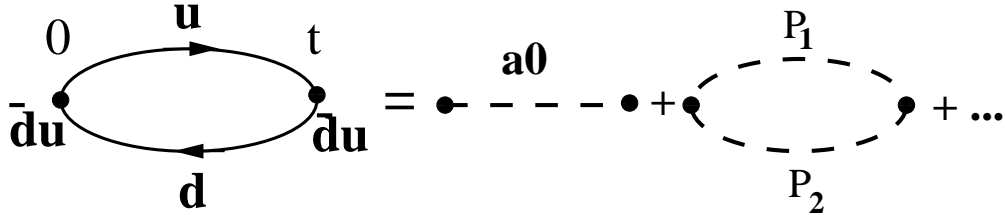,height=3cm}
\end{center}

\vspace{-0.8cm}

\caption{ \small The scalar correlator with $I=1$ receives the 
contribution from the propagation of the physical scalar meson $a_0$ 
 as well the propagation of multi-hadron states. 
The most important multi-hadron state is the two-pseudoscalar 
intermediate state, which gives rise to the so-called bubble contribution. 
 }\label{fig.cor}
\end{figure}

\begin{figure}[htb!]
\begin{center}
\epsfig{file=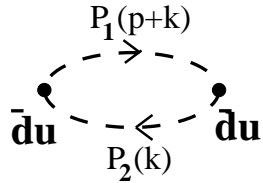,height=3cm}
\end{center}

\vspace{-0.8cm}

\caption{ \small The bubble contribution to the scalar correlator. 
Here $P_1$ and $P_2$ denote pseudoscalar fields in the relevant 
version of ChPT. }\label{fig.bubble}
\end{figure}

The issue could be settled if the mass of the lightest $\bar qq$ 
state with $I=1$ would be reliably determined on the lattice. 
We refer to this resonance  as $a_0$ and 
we consider $\bar du$ scalar meson for 
concreteness. In order to determine $a_0$ mass, lattice simulations 
evaluate the connected scalar correlator 
\begin{equation}
\label{cor}
C(t)=\sum _{\vec x}\langle 0|\bar d (\vec x,t) u(\vec x,t)  
~\bar u(\vec 0,0)d(\vec 0,0)|0\rangle~.
\end{equation}
If  $a_0$ is the lightest state of the Hamiltonian 
with quantum numbers $J^P=0^+$ and $I=1$, then $C(t)
\propto e^{-m_{a0}t}$ at large $t$ 
and extraction of $m_{a0}$ 
is straightforward. Multi-hadron states with $J^P=0^+$ and $I=1$ 
also propagate between the source and 
the sink of the scalar correlator in Fig. \ref{fig.cor} 
and they may shadow the interesting part $e^{-m_{a0}t}$. 
The most important multi-hadron state is 
 {\it the intermediate state with two pseudoscalars} $P_1P_2~$, we call 
it {\it the bubble contribution} $B$ and display it in 
Fig. \ref{fig.bubble}.   
The state $P_1P_2$ has discrete energies $E_n$ 
on the lattice with spatial extent $L$
\begin{align}
M_{P_1}+M_{P_2},\ \sqrt{M_{P_1}^2+|\tfrac{2\pi}{L}|^2}+  
\sqrt{M_{P_2}^2+|\tfrac{2\pi}{L}|^2},\ ...,
\sqrt{M_{P_1}^2+|\tfrac{2\pi}{L}\vec n|^2}+  
\sqrt{M_{P_2}^2+|\tfrac{2\pi}{L}\vec n|^2}
\end{align}
 and gives rise to a sum of decaying exponentials 
$e^{-E_nt}$ in the 
correlation function. If the masses of  
$P_1$ and $P_2$ are small, 
the bubble $B$ presents a sizable contribution  to the 
scalar correlator (Fig. \ref{fig.cor}) 
 and it has to be incorporated  in the fit of the 
lattice correlator (\ref{cor})
\begin{equation}
\label{Ctot}
C(t)=Ae^{-m_{a0}t}+B(t)+\cdots~.
\end{equation}
The dots represent contributions of excited physical scalar meson and 
other multi-hadron intermediate states, which are both less important. 

Let us review the importance 
of two-pseudoscalar intermediate state in theories with various numbers 
of dynamical quarks:
\begin{itemize}
\item The interest of this paper is $2+1$ {\it flavor QCD} ($m_u\!=\!m_d\!\not=\!m_s$), where the 
 intermediate states are $\pi\eta$, $K\bar K$ and 
$\pi \eta^\prime$ ($\pi\pi$ is not allowed by Bose symmetry 
and conservation of $J^P$ and $I^G$ in the strong decay). In Nature,
 all these intermediate states 
are lighter than $a_0(1450)$, whereas $\pi\eta$ 
is also lighter than $a_0(980)$. Hence, two-pseudoscalar 
intermediate states dominate 
correlation function at large $t$ 
if simulations are performed at small $u/d$ masses.
This makes it difficult to extract the mass of $a_0$ from (\ref{Ctot}).
For the same reason, the scalar correlator is very sensitive 
to the possible unphysical approximations undertaken in the simulations, 
like the application  of the staggered fermions or mixed quark actions.
In the present paper we calculate the bubble contribution
for the case of simulations with staggered sea and staggered valence quarks, 
as well as for simulations with staggered sea and chiral valence quarks. 
The bubble contribution incorporates the physical intermediate states 
$\pi\eta$, $K\bar K$, $\pi \eta^\prime$. It also gives the size of 
the unphysical effects of  staggered fermions and mixed quark actions in the scalar correlator. We find that these unphysical effects are sizable. 
Our results are relevant for the 
existing simulations of scalar mesons with staggered sea and valence quarks 
by MILC \cite{milc} and UKQCD \cite{irving}. They are also relevant for the 
simulations with staggered sea quarks and chiral valence quarks, currently 
performed by LHPC, NPLQCD and UKQCD  \cite{mixed_lat}. 

\item {\it Two flavor QCD} allows only 
$\pi\eta^\prime$ intermediate state, where 
$\eta^\prime$ is two-flavor singlet with mass $M_{\eta^\prime}^2=M_\pi^2+\tfrac{2}{3}m_0^2$. The $\pi\eta^\prime$ 
 state is relatively heavy and does not shadow 
so significantly the interesting part of the correlation function  given by $e^{-m_{a0}t}$. The mass of $a_0$ was extracted from the 
conventional exponential fit in the 
simulation with two flavors of dynamical Domain Wall Fermions \cite{sasa_pq}. 
The resulting mass was unaffected if the $\pi\eta^\prime$ 
intermediate state was also taken into  account in the analysis of the 
correlation function \cite{sasa_pq}. The SCALAR \cite{scalar_coll} and UKQCD 
\cite{michael} Collaborations also simulated $a_0$ in two-flavor QCD. 
Partially quenched scalar correlator with 
two dynamical Domain Wall Fermions was simulated in   \cite{sasa_pq}. 
The prediction of the bubble contribution within Partially Quenched ChPT 
describes well the striking effect of partial quenching,  
and the scalar meson mass was extracted by fitting the partially quenched 
correlators to (\ref{Ctot}) \cite{sasa_pq}.

\item The scalar correlator in {\it quenched QCD} is dominated by 
$\pi\eta^\prime$ intermediate state at light quark masses  \cite{bardeen}. 
This has 
large unphysical effect which can be attributed to the breakdown of 
unitarity in quenched QCD. The scalar correlator is negative 
and has the effective mass $2M_\pi$ at large $t$. 
The scalar meson mass $m_{a0}$ has been determined by 
fitting  
the quenched scalar correlator to the sum of 
$a_0$ exchange and 
$\pi\eta^\prime$ 
exchange\footnote{The quenched result of \cite{sasa_quenched} 
agrees with two-flavor dynamical 
result \cite{sasa_pq} if the effect of quenching 
 is incorporated at the leading order 
in the chiral expansion. The quenched $m_{a0}$ from \cite{sasa_quenched} 
is somewhat lower since the quenching effect is incorporated at 
the next-to-leading order, given by the diagrams in Fig. 8 of \cite{bardeen}.}
 (\ref{Ctot})   in \cite{bardeen,bardeen2,sasa_quenched}.
\end{itemize}

So the bubble contribution (Fig. \ref{fig.bubble}) 
to the scalar correlator is sizable and has to be incorporated 
in the fit of the lattice correlator at light quark masses.  
Our analytic prediction for the bubble contribution $B(t)$ 
offers a way to determine the scalar meson mass $m_{a0}$ by fitting 
the lattice scalar correlator to Eq. (\ref{Ctot}). 
The extraction of $m_{a0}$ in this way is possible for the range of quark 
masses and times, where $e^{-m_{a0}t}$ is not negligible with respect to the bubble contribution\footnote{The bubble contribution completely dominates the correlator for very small quark masses and large times.}.   We address 
simulations with 2+1 staggered sea quarks and staggered or 
chiral valence quarks: 
\begin{itemize}
\item 
In the case of staggered valence quarks, 
we determine the bubble contribution  
using Staggered Chiral Perturbation Theory (SChPT) \cite{schpt}. 
All the parameters entering our prediction, 
have been determined by MILC simulations. We find positive 
bubble contribution   
with effective mass $2M_{\pi_5}<m_{eff}<M_{\pi_5}+M_{\eta_5}$ at large $t$. 
This offers a natural explanation why the MILC \cite{milc} 
and UKQCD \cite{irving} Collaborations 
observed the effective mass well below 
$M_{\pi_5}+M_{\eta_5}$, which would be the lightest 
mass in proper 2+1 QCD. We propose that this unphysical result 
is due to the taste breaking.

\item  In case of chiral (Ginsparg-Wilson) valence quarks we apply 
Mixed Chiral Perturbation Theory (MChPT) \cite{mchpt}. 
The result is relevant for simulations of LHPC, NPLQCD and UKQCD  
\cite{mixed_lat} which employ staggered sea and chiral valence quarks.
The bubble contribution depends on one unknown parameter 
$a^2\Delta_{Mix}$, which gives magnitude of the taste breaking 
in the mass of the pion composed of one sea and one valence 
quark: $M_{val,sea}^2=B_0(m_{val}+m_{sea})+a^2\Delta_{Mix}$.
We find that the use of the mixed quark actions together with the 
taste breaking can have sizable unphysical effects on the scalar correlator 
and they can make the scalar correlator negative. Namely, the   
bubble contribution  is found to be negative for $a^4\Delta_{Mix}\gtrsim -0.01$ if sea and valence quark masses are tuned by imposing 
$M_{val,val}=M_{\pi_5}$ (applied by LHPC Collaboration \cite{mixed_lat}). 
The bubble contribution is positive for the reasonable values of 
$a^4\Delta_{Mix}$ if $M_{val,val}=M_{\pi_I}$ is imposed. 
\end{itemize}

\vspace{0.1cm}

The bubble contribution to the scalar correlator is calculated in Section 2. It is evaluated using Staggered  and Mixed ChPT in subsection 2.1 and 2.2, respectively. The implications of our analytical results for the  lattice simulations are also discussed in this section.  Conclusions are  given in Section 3.

\section{Contribution of two-pseudoscalar intermediate state}

In this section we calculate the contribution of two-pseudoscalar 
intermediate state 
to the scalar correlator with a point source and a point sink (\ref{cor}). 
Our calculation is carried out at the lowest order of the effective 
field theory and the relevant bubble 
diagram is shown in Fig. \ref{fig.bubble}. 
For this purpose we only need the coupling of the scalar 
current to a pair of pseudoscalar mesons $P_1P_2$ 
and the propagators for pseudoscalars $P_1$ and $P_2$.

The point scalar current is expressed in terms 
of two pseudoscalar fields $\Phi$ using the  Chiral Perturbation 
Theory \cite{bardeen}\footnote{$B_0$ was denoted by $r_0$ in 
\cite{bardeen} and by $2\mu_0$ in \cite{sasa_pq}.}
 in the Appendix A:
\begin{equation}
\label{current}
\bar  d(x)u(x) \sim B_0 [\Phi(x)^2]_{ud}~
\end{equation}  
So the coupling of the point scalar current to two pseudoscalars 
is equal to the slope 
parameter $B_0=M_\pi^2/(2m_q)$ which can be determined from the lattice 
data. The same coupling applies also 
 in  the Staggered as well as in the  Mixed version of ChPT 
at the lowest order. 

The bubble contribution in Fig. \ref{fig.bubble} is equal to 
\begin{equation}
\label{B}
B=\langle 0 | \bar d u~\bar u d|0\rangle_{bubble}=B_0^2\langle 0 | [\Phi^2]_{ud}[\Phi^2]_{du}|0\rangle~.
\end{equation}
The Wick contractions 
lead to the products of two propagators for pseudoscalar fields. 
The propagators and the resulting 
bubble contribution will be given separately for 
Mixed and Staggered ChPT in two subsections 
that follow.

We note that the bubble contribution (\ref{B}) applies 
for the bare lattice  point-point  correlator (\ref{cor}) via (\ref{Ctot}), 
if the bare quark mass is 
inserted to $B_0=M_\pi^2/(2m_q)$. This is possible since 
the scalar current and $B_0\propto 1/m_q$ are multiplied by the same factor 
$Z_S$ in the renormalization.  
 
\subsection{Staggered sea quarks and staggered valence quarks}

First we consider the simulations with staggered valence quarks 
and three flavors of staggered sea quarks with $m_u=m_d\not =m_s$. 
We study only the dynamical correlators, where the valence-quark mass 
is equal to the $u/d$ sea-quark mass. 
Staggered quarks have four tastes and 
the scalar correlator is conventionally 
calculated using the taste-singlet source and sink. 
This correlator receives contribution from the physical scalar meson and 
the bubble contribution (\ref{Ctot}). It also receives an oscillating 
contribution from a pseudoscalar meson\footnote{The operator with spin $I$ and taste $I$ couples also to state with spin $\gamma_5\gamma_4$ and taste $\gamma_5\gamma_4$.}, but the oscillating contribution will not  
be considered here since it  has  been    
identified and removed in the analysis of the lattice data \cite{milc,irving}. 

The bubble contribution (\ref{B}) in a theory with 
four tastes and three flavors of sea quarks ($4+4+4$) is equal to 
\begin{equation}
\label{schpt_0}
B_{4+4+4}^{SChPT}=\tfrac{1}{4}B_0^2 \sum_{t,t^\prime=1}^{4}\langle 0|[\Phi^2]_{ut~dt}[\Phi^2]_{dt^\prime~ ut^\prime}|0\rangle~,
\end{equation}
where $1/4$ is appropriate normalisation. 
Here $\Phi$ is $12\times 12$ pseudoscalar matrix of Staggered 
ChPT \cite{schpt} and the subscripts  
$u,d$ denote its flavor component, whereas
  $t,t^\prime=1,..,4$ denote its taste 
component. The field $\Phi=\sum_{b=1}^{16}\tfrac{1}{2}T^b\phi^b$ 
is expressed in terms of mass eigenstates $\phi^b$, where $\phi^b$ 
is $3\times 3$ pseudoscalar matrix with flavor components 
$\phi^b_{ff^\prime}$ and 
$T^b=\{\xi_5,i\xi_5\xi_\mu,i\xi_\mu\xi_\nu,\xi_\mu,\xi_I\}$. The bubble contribution can be expressed in terms of the pseudoscalar propagators  $\langle \phi^b|\phi^b\rangle$ by applying the Wick contractions to (\ref{schpt_0}) 
\begin{equation}
\label{schpt_1}
B^{SChPT}_{4+4+4}
=\tfrac{1}{4}B_0^2\sum_{b=1}^{16}\biggl(\langle \phi^b_{us}|\phi^b_{su}\rangle \langle \phi^b_{sd}|\phi^b_{ds}\rangle
+\langle \phi^b_{ud}|\phi^b_{du}\rangle [\langle \phi^b_{uu}|\phi^b_{uu}\rangle+\langle \phi^b_{dd}|\phi^b_{dd}\rangle+2\langle \phi^b_{uu}|\phi^b_{dd}\rangle]\biggr)~.
\end{equation}

\begin{figure}[htb!]
\begin{center}
\epsfig{file=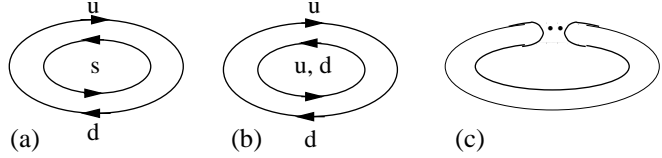,height=2cm}
\end{center}

\vspace{-0.8cm}

\caption{ \small The quark level diagrams that correspond 
to the bubble contribution (Fig. \ref{fig.bubble}) in simulations with 
 dynamical $u$, $d$ and $s$ quarks. The diagram (c) represents 
also the diagrams with any number of
 quark loops inserted in the place of the dots. }\label{fig.diagrams}
\end{figure}

The propagators $\langle \phi^b|\phi^b\rangle$  for the 
pseudoscalar fields of various tastes $b$ 
are provided by SChPT \cite{schpt} and are collected in the Appendix B. 
Propagators for all tastes ($I,V,A,T,P$) have connected part, while tastes 
$I,~V$ and $A$ have also disconnected part. The bubble contribution  
(\ref{schpt_1}) can be represented by the quark-level Feynman diagrams in Fig. \ref{fig.diagrams} and it is calculated by inserting the mesonic propagators  in Appendix B  to expression (\ref{schpt_1}). We get 
\begin{align}
\label{Bschpt}
B^{SChPT}(t)=F.T.[~B^{SChPT}&(p)~]_{~\vec p=\vec 0}\quad {\rm with} \\
B^{SChPT}(p)=B_0^2 \! \sum_k\Biggl\{
-4\biggl[&\frac{1}{(k+p)^2+M_{U_I}^2}~ \frac{1}{3}~
\frac{(k^2+M_{S_I}^2)}{(k^2+M_{U_I}^2)(k^2+\tfrac{1}{3}M_{U_I}^2+\tfrac{2}{3}M_{S_I}^2)}\nonumber\\
+&\frac{1}{(k+p)^2+M_{U_V}^2}~a^2\delta_V^\prime~\frac{(k^2+M_{S_V}^2)}{(k^2+M_{U_V}^2)(k^2+M_{\eta_V}^2)(k^2+M_{\eta^\prime_V}^2)}\nonumber\\
+&\frac{1}{(k+p)^2+M_{U_A}^2}~a^2\delta_A^\prime~\frac{(k^2+M_{S_A}^2)}{(k^2+M_{U_A}^2)(k^2+M_{\eta_A}^2)(k^2+M_{\eta^\prime_A}^2)}
\biggr]\nonumber\\
+C~\frac{1}{4}\sum_{b=1}^{16}& \biggl[2~ \frac{1}{(k+p)^2+M_{U_b}^2}~ \frac{1}{k^2+M_{U_b}^2}+\frac{1}{(k+p)^2+M_{us_{b}}^2}~ \frac{1}{k^2+M_{us_{b}}^2}\biggr]\Biggl\}~,\nonumber
\end{align}
where
\begin{align}
\label{444_111}
C&=1\ ,\ \   M_{\eta^{(\prime)}_{V,A}}=m_{\eta^{(\prime)}}(a^2\delta_{V,A}^\prime)\quad \ \  {\rm in}\  2\cdot 4+4 \ {\rm theory}\\
C&=\tfrac{1}{4}\ ,\ \   M_{\eta^{(\prime)}_{V,A}}=m_{\eta^{(\prime)}}(\tfrac{1}{4}~a^2\delta_{V,A}^\prime)\quad {\rm in} \ 2+1 \ {\rm theory}\nonumber
\end{align}
and the functions $m_{\eta}$, $m_{\eta^\prime}$ are given by  (\ref{meta})
in Appendix B. 
The four tastes per sea quark  
($2\cdot 4+4$) are reduced to one taste per sea quark 
 ($2+1$) by applying  $\root 4 \of{Det}$ trick in staggered lattice simulations \cite{milc}. In SChPT \cite{schpt} this is achieved by multiplying every sea  quark loop by factor $1/4$. The sea quark loops are present 
in diagrams (a,b) and in hairpin vertex of diagram (c) in Fig. \ref{fig.diagrams}. The corresponding factors  of $1/4$ are incorporated in (\ref{444_111}).

The pseudoscalar masses $M_{U_b}\!\equiv\! M_{\pi_b}\!=\!2B_0m_u+a^2\Delta_b$, $M_{S_b}\!=\!2B_0m_s+a^2\Delta_b$ and  $M_{us_b}\!=\!B_0(m_u+m_s)+a^2\Delta_b$  of unmixed mesons have been determined by MILC simulations \cite{milc}. The taste breaking $a^2\Delta_b$ is flavor independent and vanishes in the continuum limit. The hairpin 
couplings of the taste-vector and taste-axial mesons were also determined 
by MILC for their coarse and fine lattices
\cite{milc3} 
\begin{equation}
\label{hairpin}
(a^4\delta^\prime_{A})_{\rm coarse}=-0.04(1)~,\quad (a^4\delta^\prime_V)_{\rm coarse}=-0.01(\textstyle{+3\atop -1})~,\quad (a^2 \delta^\prime_{A,V})_{\rm fine}\simeq 0.35~(a^2 \delta^\prime_{A,V})_{\rm coarse}~.  
\end{equation}
These hairpin couplings mix the flavor neutral mesons of tastes $V$ and $A$  
to the mass eigenstates $\eta_{V,A}$ and $\eta^\prime_{V,A}$ 
with masses given in (\ref{444_111},\ref{meta}). 

\vspace{0.4cm}

{\bf a) Results in case of 4 tastes per quark }

\vspace{0.2cm}

The theory with four tastes per sea quark (i.e. without $\root 4 \of{Det}$ trick)  is not relevant to QCD. In spite of that, it is illuminating  
to consider the properties of the bubble contribution in this case before 
proceeding to the single-taste theory. 

The four-taste theory 
is a proper unitary quantum field theory even at finite lattice spacing 
and  $B^{SChPT}_{2\cdot 4+4}(t)$ (\ref{Bschpt}) is positive definite 
for any values of the input parameters. 
This is explicitly demonstrated for the simplified case of 
$m_u\!=\!m_d\!=\!m_s$ and $\Delta_b\!=\!0$ with 
 general values of hairpin couplings 
$a^2\delta^\prime_{V,A}$ in Appendix C. 

The $B^{SChPT}_{2\cdot 4+4}(t)\propto e^{-2M_\pi t}$ 
at large time for {\it zero}  
and non-zero lattice spacing. This is in sharp contrast to three flavor QCD, 
where  $B(t)\propto e^{-(M_\pi+M_\eta) t}$ with 
$M_\eta^2=\tfrac{1}{3}M_\pi^2+\tfrac{2}{3}M_S^2$. It is easy to understand why an intermediate state with mass $2M_\pi$ is possible in $B^{SChPT}_{2\cdot 4+4}$ (\ref{Bschpt}) 
at $a=0$. The mass eigenstates that enter as intermediate states 
are $\pi_b\eta_b$ and $K_b\bar K_b$. Only the mass of $\eta_I$ is lifted 
by the anomaly to  $\sqrt{\tfrac{1}{3}M_\pi^2+\tfrac{2}{3}M_S^2}$, while the etas of 
other tastes retain mass $M_\pi$ at $a=0$. So the lightest intermediate state with $I=1$ has the mass $2M_\pi$. 
 
\vspace{0.4cm}

{\bf b) Results in case of one taste per sea quark}

\vspace{0.1cm}

Finally we study the properties of $B_{2+1}^{SChPT}$ (\ref{Bschpt}), 
which is relevant to QCD since  the four tastes of sea quarks 
have been reduced to one via (\ref{444_111}). 

In the continuum limit there is no taste breaking ($a^2\Delta_b\to 0$, $a^2\delta^\prime_{V,A}\to 0$) and $B_{2+1}^{SChPT}$ (\ref{Bschpt}) reduces to the 
contributions of $\pi\eta$ and $K\bar K$ from $2+1$ ChPT
\begin{align}
\label{Bchpt}
B^{ChPT}(p)&=B_0^2 \! \sum_k\Biggl\{
\frac{2}{3} \frac{1}{(k+p)^2+M_{\pi}^2}~
\frac{1}{k^2+M_{\eta}^2}
+\frac{1}{(k+p)^2+M_{K}^2}~ \frac{1}{k^2+M_{K}^2}\Biggl\}\nonumber\\
\lim_{t\to\infty}B^{ChPT}(t)&=\frac{B_0^2}{4 L^3}\biggl(\frac{2}{3}\frac{e^{-(M_{\pi}+M_\eta)t}}{M_{\pi}M_\eta}+\frac{e^{-2M_Kt}}{M_{K}^2}\biggr)~,
\end{align}
where $M_\eta^2=\tfrac{1}{3}M_{\pi}^2+\tfrac{2}{3}M_{S}^2$ and  
$M_K^2=\tfrac{1}{2}M_{\pi}^2+\tfrac{1}{2}M_{S}^2$
The bubble contribution is positive and  drops as 
$e^{-(M_{\pi}+M_\eta)t}$ at large $t$, as expected in QCD. 

The fourth-root trick destroys the unitarity of the theory at finite lattice spacing and $B_{2+1}^{SChPT}(t)$ can be negative for specific values of 
input parameters\footnote{For example, $B_{2+1}^{SChPT}(t)$ is negative
  in the case of 
input parameters relevant for MILC simulations with sign of $a^2\delta^\prime_{V,A}$ in (\ref{hairpin}) reversed.}. On top of that, the taste breaking 
at nonzero $a$ allows for the intermediate state with $2M_\pi$, which
is prohibited in proper QCD. This  
can be understood by considering the quark flow diagrams in 
Fig. \ref{fig.diagrams}. The diagram in Fig. \ref{fig.diagrams}a exhibits
 the expected time-dependence $e^{-2M_Kt}$ in ChPT as well as in SChPT. 
The diagrams in Figs. 
\ref{fig.diagrams}b and \ref{fig.diagrams}c both contain terms with 
$e^{-2M_\pi t}$, but these exactly cancel in ChPT 
with no taste breaking. This cancellation does not occur in the 
presence of taste breaking when pions of all tastes  
$\pi_{5,T,I,V,A}$ flow in diagram \ref{fig.diagrams}b, but only the 
pions $\pi_{I,V,A}$ flow in the diagram \ref{fig.diagrams}c. 
The taste breaking therefore significantly modifies the time-dependence 
\begin{equation}
\lim_{t\to\infty}B^{SChPT}_{2+1}(t)=\frac{B_0^2}{4 L^3}\biggl(\frac{2}{16}\frac{e^{-2M_{\pi_5}t}}{M_{\pi_5}^2}+{{\rm terms\ with}\atop e^{-2M_{\pi_{b\not =5}}}}+ {\rm {pairs\atop \pi\eta,\pi\eta^\prime,K\bar K}}\biggr)
\end{equation} 
with respect to the expected behaviour in proper QCD (\ref{Bchpt}). 
The coupling to the un-physical intermediate state with mass $2M_\pi$ 
 vanishes only in the continuum limit.  
   
\vspace{0.2cm}

\begin{figure}[htb!]
\begin{center}
\epsfig{file=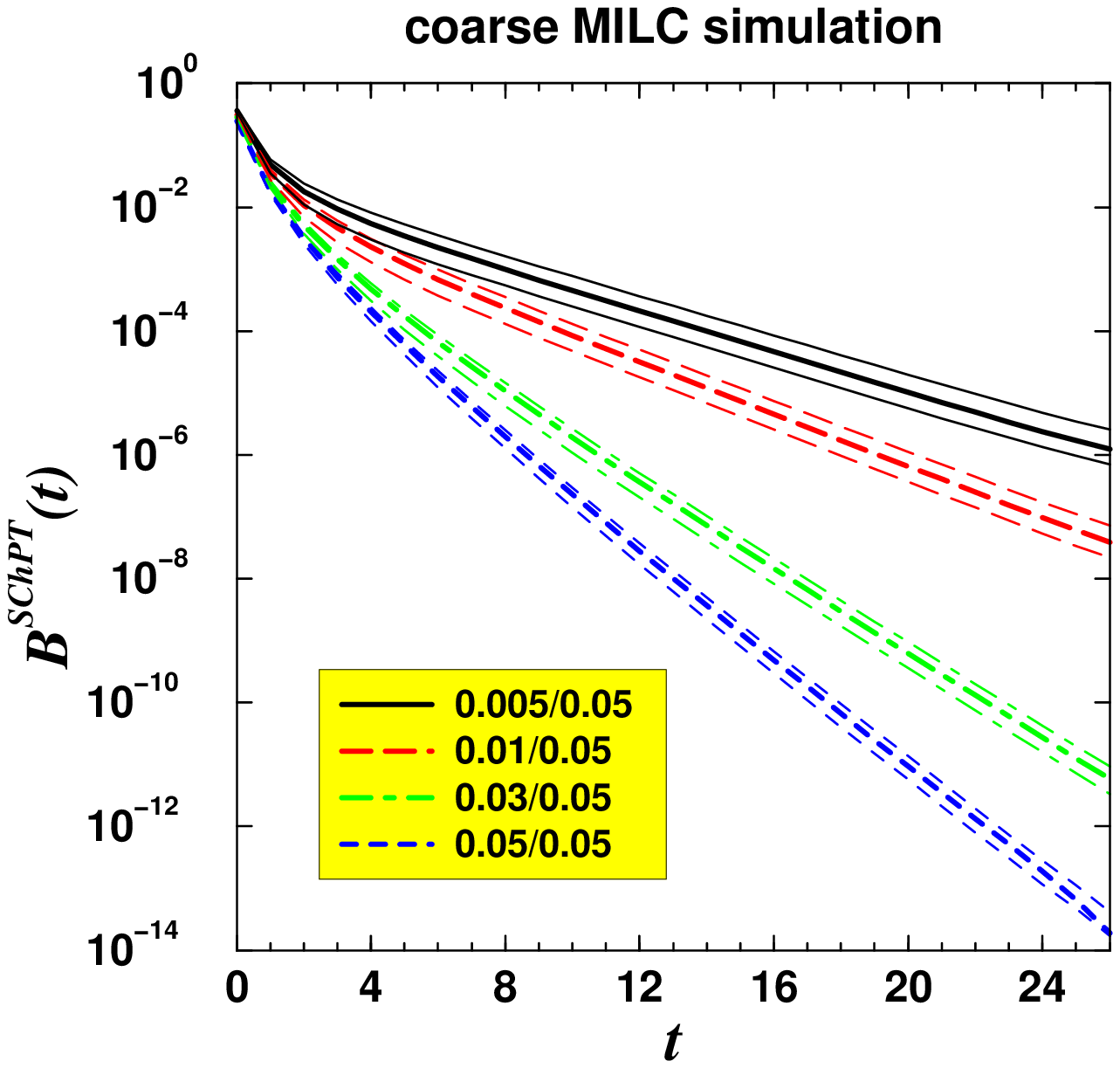,height=7cm} 
$\quad$
\epsfig{file=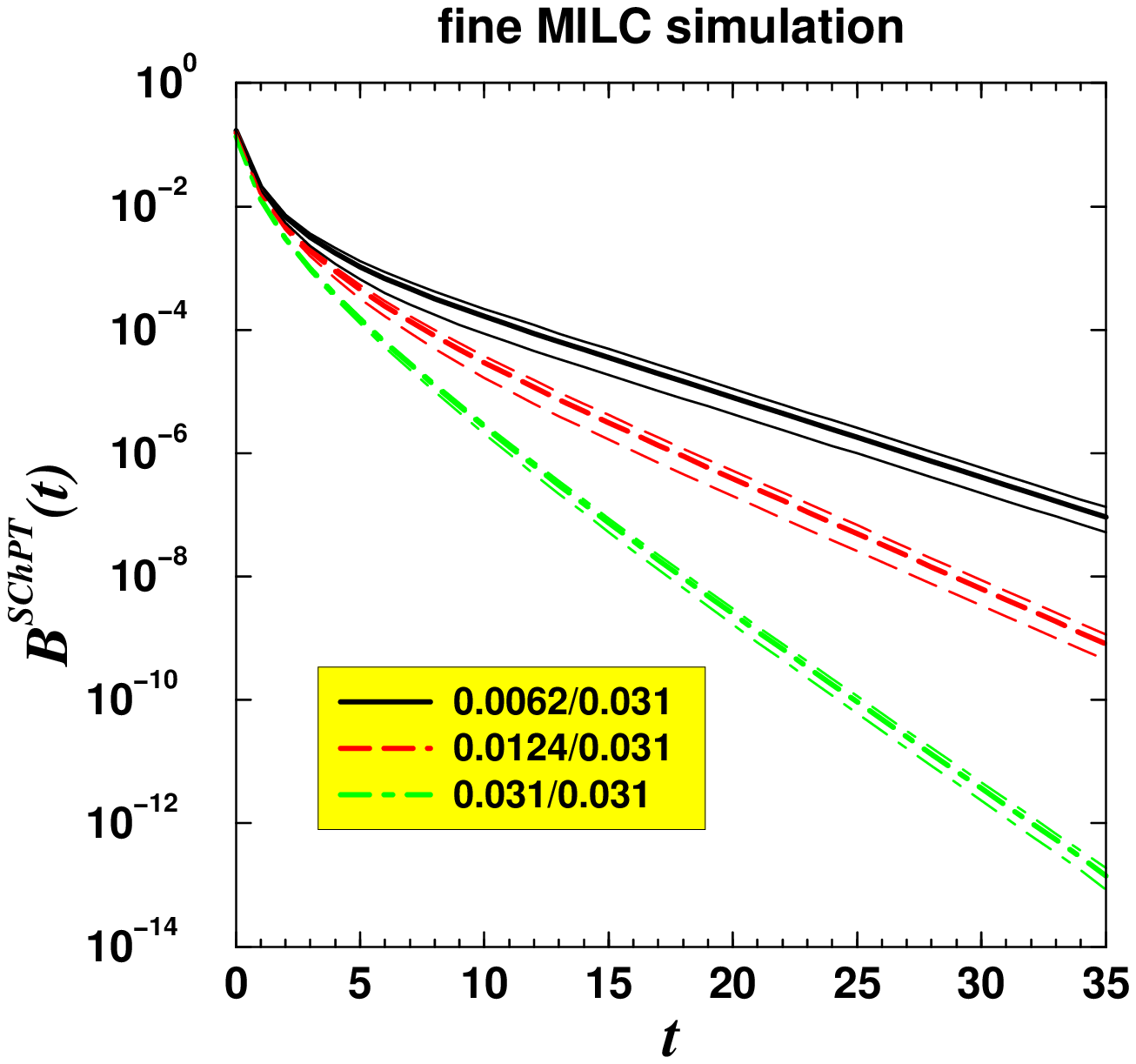,height=7cm}
\end{center}

\vspace{-0.8cm}

\caption{ \small The bubble contribution $B^{SChPT}_{2+1}(t)$ (\ref{Bschpt}) for 
MILC simulation with $2+1$ staggered sea quarks and staggered valence quarks. 
The $am_{u/d}/am_s$ denote the $2+1$ 
sea-quark 
masses in lattice units. The coarse and fine lattices have the  volumes $20^3\times 64$ and $28^3\times 96$, respectively. The thiner lines denote the variation of $B^{SChPT}_{2+1}(t)$ due to the uncertainty 
in the hairpin parameters (\ref{hairpin}). }\label{fig.schpt_cor}
\end{figure}

\begin{figure}[htb!]
\begin{center}
\epsfig{file=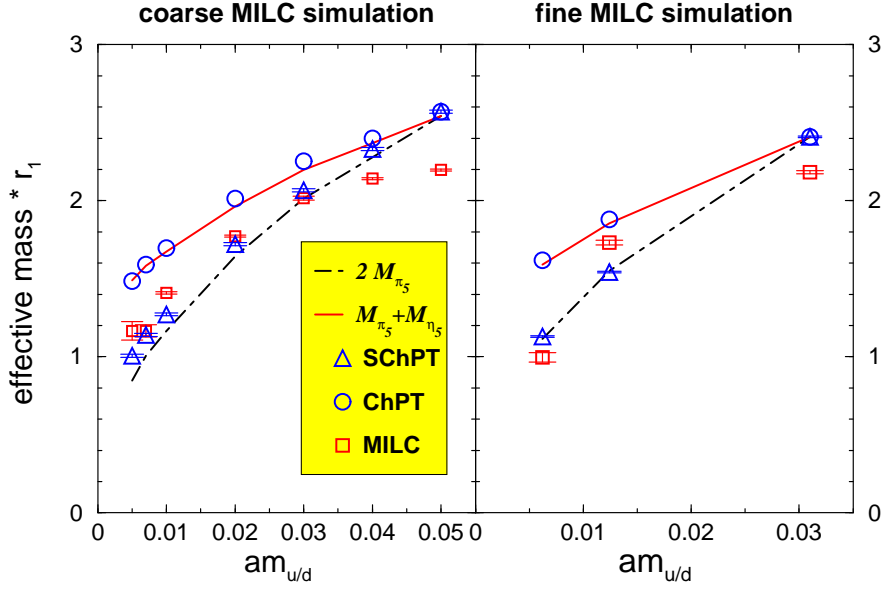,height=8cm}
\end{center}

\vspace{-0.8cm}

\caption{ \small The effective mass of the scalar correlator 
for MILC simulation with $2+1$ staggered sea quarks and staggered valence quarks. The effective mass is multiplied by $r_1\simeq 1.6$ GeV$^{-1}$
 as in \cite{milc} for easier comparison. The dependence on $am_{u/d}$ is 
shown, while $am_s$ is fixed to $am_s=0.05$ and $am_s=0.031$ on coarse and 
fine lattices, respectively.   Triangles present effective 
mass of the bubble contribution in $2+1$ Staggered ChPT 
(\ref{Bschpt}) 
at $t=20$ for coarse MILC lattice and at $t=30$ for fine MILC lattice; 
the corresponding 
error-bar denotes the variation due to the uncertainty in the hairpin parameters (\ref{hairpin}). 
In the continuum limit  $B^{SChPT}_{2+1}$ (\ref{Bschpt}) reduces to 
$B^{ChPT}$ (\ref{Bchpt}) and its effective mass 
is given by circles. Squares represent effective mass 
from MILC simulation \cite{milc}. The relevant masses $2M_{\pi_5}$ and 
$M_{\pi_5}+M_{\eta_5}$ from MILC simulations are also shown 
($M_{\eta_5}^2\equiv \tfrac{1}{3}M_{U_5}^2+\tfrac{2}{3}M_{S_5}^2$).
}\label{fig.schpt_eff}
\end{figure}

Now we evaluate $B_{2+1}^{SChPT}$ (\ref{Bschpt}) for the case of 
MILC configurations in order to see if our observations 
agree with the results from the lattice simulations.  Fig. \ref{fig.schpt_cor}
shows the magnitude of the bubble contribution  evaluated 
for the $2+1$  MILC coarse and fine lattices, using the 
pseudoscalar masses, the bare quark masses, the hairpins (\ref{hairpin}) and 
the lattice volumes  from \cite{milc}.  The Euclidean loop momentum 
$k$ is summed over the discrete  momenta in the finite box of the lattice\footnote{The detailed expression for the finite sum in $B(t)$ is given in Appendix B of \cite{sasa_pq} for analogous case of Partially Quenched theory.}. 
The bubble contribution is larger for 
lighter $u/d$ quark masses. It becomes the dominant contribution 
to the scalar correlator at large $t$ if  two-pseudoscalar state 
is lighter than $a_0$, which is realized\footnote{
This applies for observed $m_{a0}=1450$ MeV as well as $m_{a0}$ obtained 
by MILC from heavier quark masses.}  
for the $u/d$ quark masses considered in Figs.  \ref{fig.schpt_cor} and 
 \ref{fig.schpt_eff}.  

The effective mass of $B^{SChPT}_{2+1}(t)$ 
is shown by triangles in  Fig. \ref{fig.schpt_eff}. 
It is compared with $M_{\pi_5}+M_{\eta_5}$, $2M_{\pi_5}$ and with 
 the effective mass of the scalar correlator from 
MILC simulation \cite{milc}. The effective mass of the analytic prediction 
 lies between 
$2M_{\pi_5}< m_{eff}< M_{\pi_5}+M_{\eta_5}$ 
at large $t$. This agrees\footnote{Fig. \ref{fig.schpt_eff} presents  
the effective mass of $B^{SChPT}_{2+1}$ at $t=20$  
for coarse and at $t=30$ for fine lattice, 
while the MILC effective mass was extracted 
from a wider time-range $t\sim 5-20$ \cite{milc}, so they are not expected to agree exactly.}
  with the effective mass observed in 
lattice simulations by MILC \cite{milc} (shown in Fig. \ref{fig.schpt_eff})  
and by UKQCD \cite{irving}.
The exception are the highest 
quark masses,  where the lattice simulations \cite{milc,irving} 
observe effective mass even below $2M_{\pi_5}$, which can not be understood 
within given theoretical framework. The unphysical effective mass 
$2M_{\pi}$ is lifted in the continuum limit to the physical value 
$M_{\pi}+M_{\eta}$, which is the effective mass of  $B^{ChPT}$ 
(\ref{Bchpt})  and is represented by circles in Fig. \ref{fig.schpt_eff}.  

\subsection{Staggered sea quarks and chiral valence quarks} 

The simulations with  chiral valence quarks on the available staggered MILC configurations present an appealing possibility.  The LHPC and NPLQCD Collaborations  use domain wall valence quarks, while UKQCD uses overlap valence quarks \cite{mixed_lat}. The corresponding effective theory has been derived recently 
\cite{mchpt} and we refer to it as Mixed ChPT (MChPT). It is a kind of 
Partially Quenched theory which takes into account also the 
taste breaking of the staggered sea quarks. The valence quarks $x=1,2$ and the ghost valence quarks $\tilde x=\tilde 1,\tilde 2$ have degenerate 
mass $m_{val}$, while the staggered sea quarks $4u,4d,4s$ have masses $m_u=m_d$ and $m_s$. There is no unique recipe how to match the valence quark mass $m_{val}$ and the sea quark mass $m_{u}$, so the mixed theory always retains some features of partial quenching away from the continuum limit. 

The  bubble contribution is obtained by applying the Wick contractions to (\ref{B})
\begin{equation}
\label{mchpt_1}
 B^{{\small MChPT}}_{4+4+4}\!=\!B_0^2\biggl[ 2\langle\Phi_{11}|\Phi_{22}\rangle\langle\Phi_{12}|\Phi_{21}\rangle
+\!\!\!\!\sum_{x=1,2,\tilde 1,\tilde 2}\!\!\!\langle\Phi_{1x}|\Phi_{x 1}\rangle \langle\Phi_{x 2}|\Phi_{2 x}\rangle+\!\!\!\!\!\!\!\!\!\!\sum_{sea=4u,4d,4s}\!\!\!\!\!\!\langle\Phi_{1~\!sea}|\Phi_{sea~ \!1}\rangle\langle\Phi_{sea~\!2}|\Phi_{2~\!sea}\rangle\biggr].
\end{equation}
To reduce four tastes per sea quark to one taste, the last term will be multiplied by $1/4$ and  the appropriate expressions for disconnected propagators from \cite{mchpt} will be used.

The propagators for valence-valence mesons with $x,x^\prime=1,2,\tilde 1,\tilde 2$ and $m_0\to\infty$ are \cite{mchpt}   
\begin{align}
\label{mchpt_prop_vv}
&\langle\Phi_{xx^\prime}|\Phi_{x^\prime x}\rangle=\frac{\epsilon_x}{k^2+M_{val,val}^2}~,\quad M_{val,val}^2=2B_0 m_{val}~,\quad \epsilon_{1,2}=1~,\ \epsilon_{\tilde 1,\tilde 2}=-1\\
&\langle\Phi_{xx}|\Phi_{x^\prime x^\prime}\rangle_{disc}=-\frac{1}{3}\frac{(k^2+M_{U_I}^2)(k^2+M_{S_I}^2)}{(k^2+M_{val,val}^2)^2(k^2+\tfrac{1}{3}M_{U_I}^2+\tfrac{2}{3}M_{S_I}^2)}~,\quad {\small {M_{U_I}^2=M_{U_5}^2+a^2\Delta_I\atop M_{S_I}^2=M_{S_5}^2+a^2\Delta_I}}~,\nonumber
\end{align}
while the propagators for valence-sea mesons with $sea=u,d,s$ and $x=1,2$ are 
\begin{equation}
\label{mchpt_prop_vs}
\langle\Phi_{x~\!sea}|\Phi_{sea~\! x}\rangle=\frac{1}{k^2+M_{val,sea}^2}~,\quad  M_{val,sea}^2=B_0(m_{val}+m_{sea})+a^2\Delta_{Mix}~.
\end{equation}
The propagators (\ref{mchpt_prop_vv},\ref{mchpt_prop_vs}) 
depend only on two taste breaking parameters $a^2\Delta_I$ and 
$a^2\Delta_{Mix}$. The taste-singlet breaking $a^2\Delta_I$ of the  
sea-sea pion mass is taken from \cite{milc}, while  the 
taste breaking $a^2\Delta_{Mix}$  of the valence-sea pion mass 
has not been determined yet and it is the only free parameter 
in  the present paper. 

The bubble contribution is obtained by inserting the propagators (\ref{mchpt_prop_vv},\ref{mchpt_prop_vs}) to (\ref{mchpt_1})
\begin{align}
\label{Bmchpt}
B_{2+1}^{MChPT}(t)&=F.T.[~B_{2+1}^{MChPT}(p)~]_{~\vec p=\vec 0}\quad {\rm with}\\
B_{2+1}^{MChPT}(p)&=B_0^2 \! \sum_k\Biggl\{-\frac{4}{3}~\frac{1}{(k+p)^2+M_{val,val}^2}~\frac{1}{(k^2+M_{val,val}^2)^2}~
\frac{(k^2+M_{U_I}^2)(k^2+M_{S_I}^2)}{k^2+\tfrac{1}{3}M_{U_I}^2+\tfrac{2}{3}M_{S_I}^2}\nonumber\\
&\ \ \ \ \ \ \ \ \ \ \ \ +2~ \frac{1}{(k+p)^2+M_{val,u}^2}~ \frac{1}{k^2+M_{val,u}^2}+\frac{1}{(k+p)^2+M_{val,s}^2}~ \frac{1}{k^2+M_{val,s}^2}\Biggl\}~.\nonumber
\end{align}
Our result agrees in appropriate 
limit\footnote{The results agree in the limit 
$t\to\infty$, $a^2\Delta_{Mix}\to 0$ and $m_u=m_d=m_s$.}  
with the scalar correlator 
from Ref. \cite{mchpt_taku}, which considers mixed quark actions 
but not staggered sea.  In  the continuum limit 
($a^2\Delta_I\to 0$, $a^2\Delta_{Mix}\to 0$) the expression  (\ref{Bmchpt}) 
reduces to $B^{ChPT}$ (\ref{Bchpt}) once the valence and the 
sea quark masses are tuned as $m_{val}=m_u$. Away from the continuum limit,
the contributions $e^{-2M_\pi t}$ from  
the diagrams in Figs. \ref{fig.diagrams}b and \ref{fig.diagrams}c
do not cancel  and the correlator has the effective mass $2M_\pi$ 
at large $t$. 

\begin{figure}[htb!]
\begin{center}
\epsfig{file=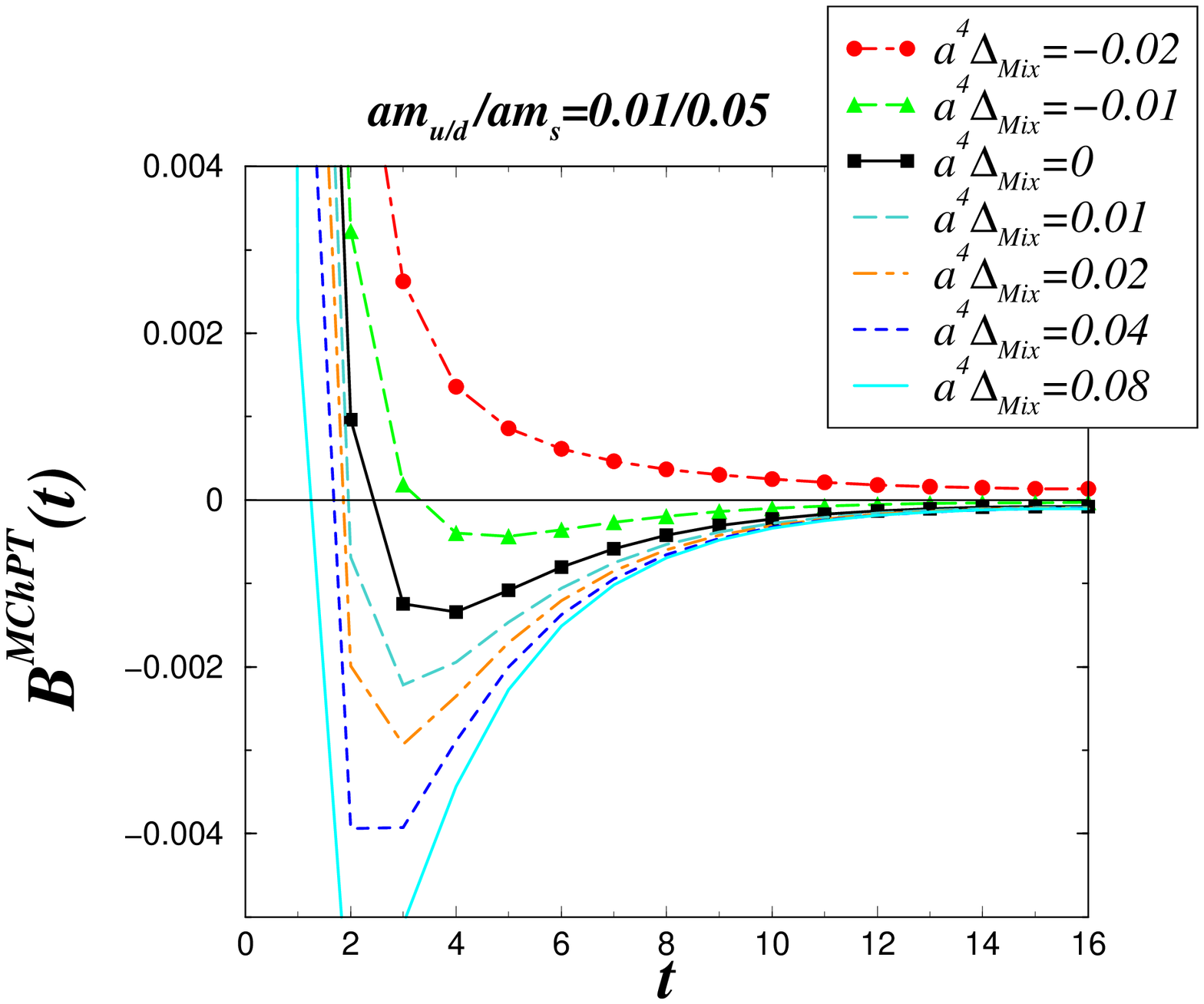,height=8cm}
$\qquad\qquad$
\epsfig{file=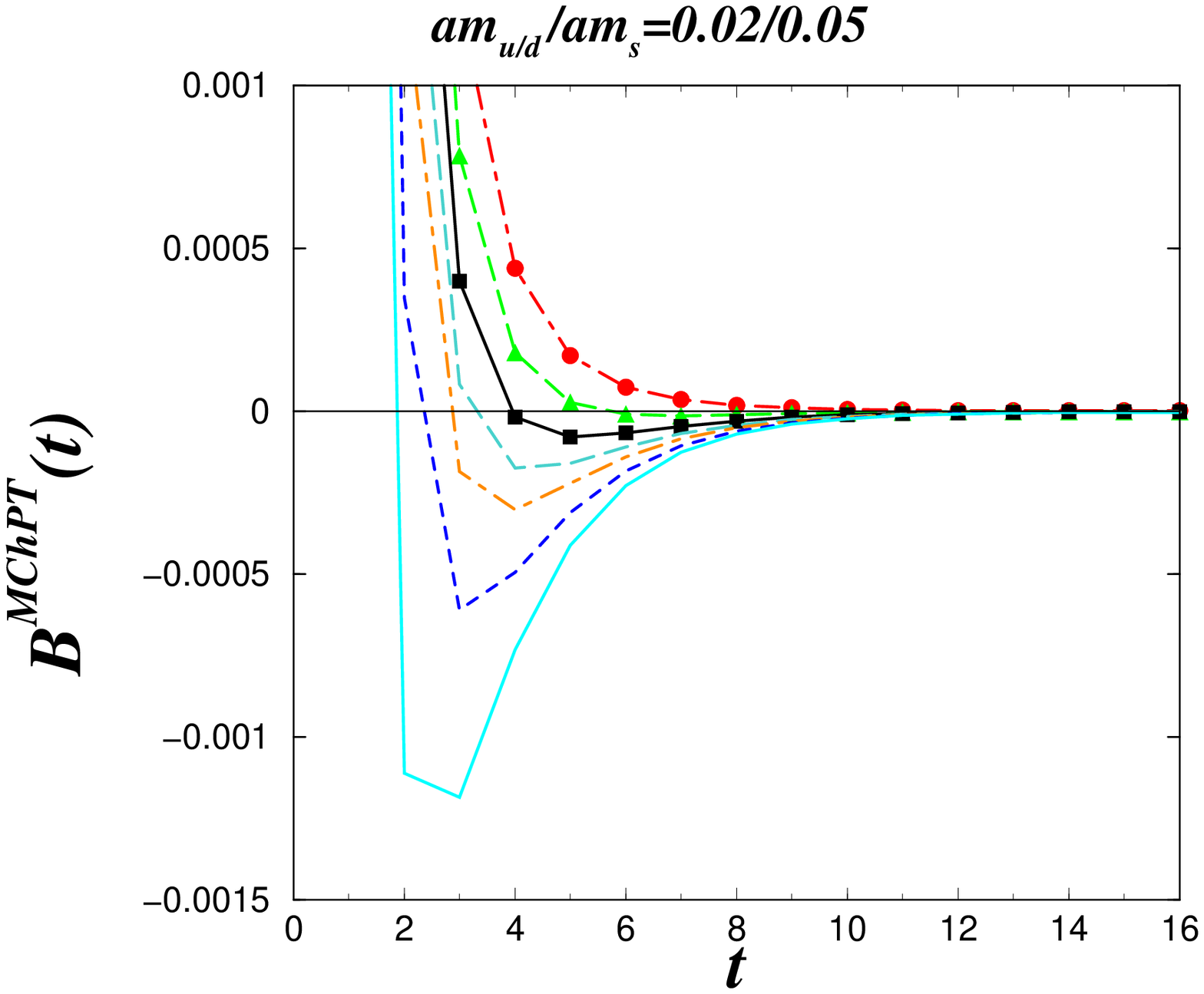,height=7cm}$\quad$
\epsfig{file=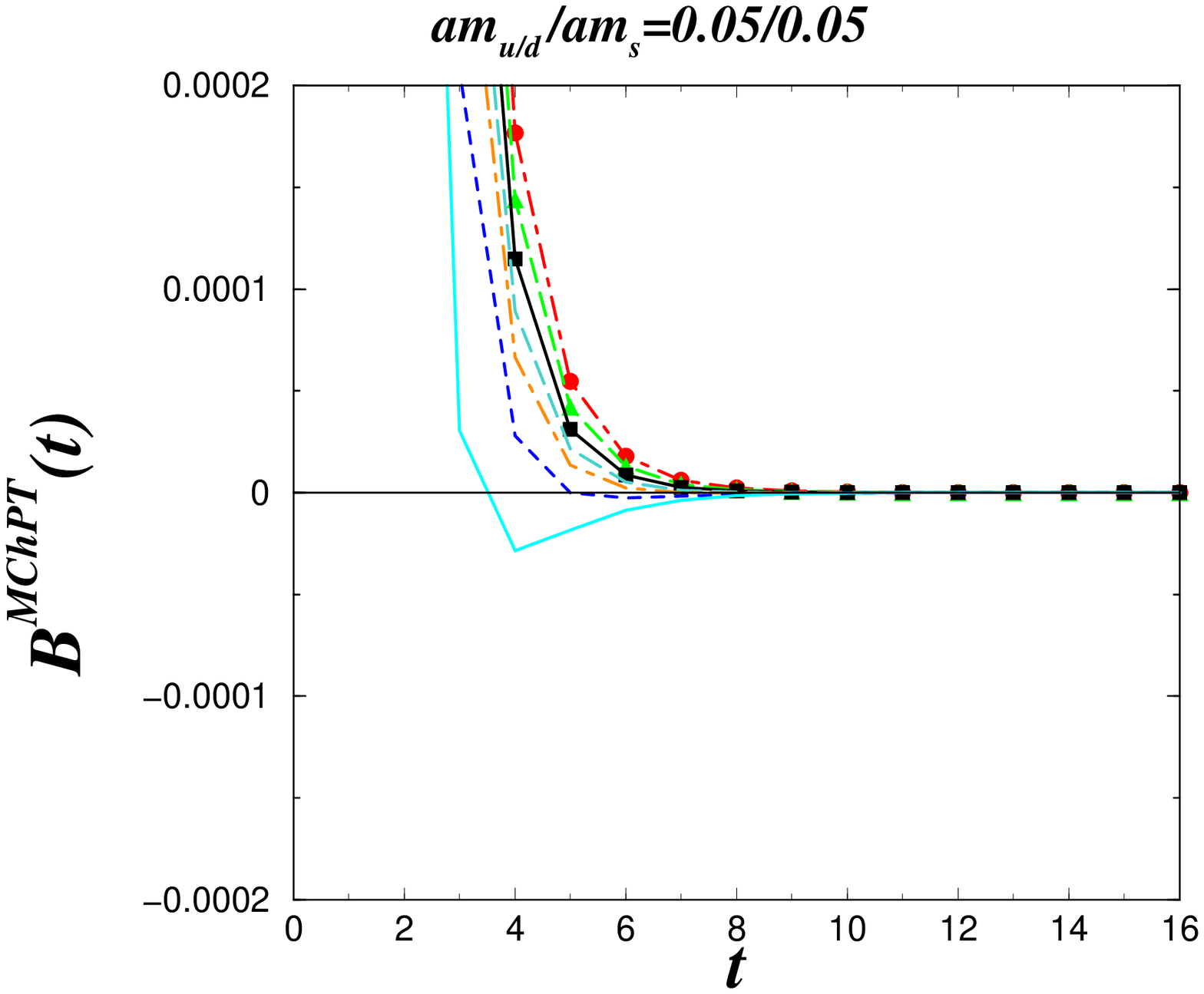,height=7cm}
\end{center}

\vspace{-0.8cm}

\caption{ \small The bubble contribution $B_{2+1}^{MChPT}(t)$ 
(\ref{Bmchpt}) for 
the simulation with chiral fermions on $2+1$ MILC staggered 
configurations with  volume  $20^3\times 32$ and coarse $a$.  The valence and sea quark masses are tuned 
by matching $M_{val,val}=M_{U_5}$. The unknown parameter $a^4\Delta_{Mix}$ 
is varied in the reasonable range  $-0.02\leq a^4\Delta_{Mix}\leq 0.08$.}\label{fig.mchpt_tune1}
\end{figure}

Fig. \ref{fig.mchpt_tune1} represents the bubble contribution (\ref{Bmchpt}) 
 when valence and sea quark masses are tuned by matching 
$M_{val,val}=M_{U_5}$ which is used  by LHPC 
Collaboration \cite{mixed_lat}. The expression (\ref{Bmchpt}) is evaluated for MILC coarse lattices 
using $M_{U_5}$ and $V=20^3\times 32$ from LHPC \cite{mixed_lat} 
and $a^2\Delta_I$  from \cite{milc}. 
The $a^2\Delta_{Mix}$ is the only unknown 
parameter and it is varied in the reasonable\footnote{The natural limit 
for the magnitude of taste breaking is $|a^4\Delta_{Mix}|<a^4\Delta_I\simeq 0.08$. The additional restriction in case of negative $a^2\Delta_{Mix}$ 
is the positivity of the mass $M_{val,sea}$, which requires 
$a^4\Delta_{Mix}\gtrsim -0.03$ 
for the simulation with $am_{u/d}/am_s=0.007/0.05$. } 
range $-0.02\leq a^4\Delta_{Mix}\leq 0.08$.  The bubble contribution 
is sizable for small $m_{u/d}$ and we find it is negative for 
$a^4\Delta_{Mix}\gtrsim -0.01$. 
This negativity 
can be attributed to the fact that 
all the sea-sea pions (except $\pi_5$) 
are heavier than the valence-valence pion
 when $M_{val,val}=M_{U_5}$ 
and the correlator goes negative for the same reason as in Partially 
Quenched QCD \cite{sasa_pq}. At small $m_{u/d}$,  
the scalar correlator (\ref{Ctot}) is dominated 
by the bubble contribution and is therefore negative 
for a large range of  values $a^2\Delta_{Mix}$. 
At large $m_{u/d}$, the scalar correlator is dominated by 
$e^{-m_{a0}t}$ for moderate $t$, 
so it is positive irrespective 
of the sign of the bubble contribution. 
A lattice study of point-point scalar correlator at small quark  masses   
offers a way to determine the unknown parameter $a^2\Delta_{Mix}$ 
via Eq. (\ref{Bmchpt}). 

\begin{figure}[htb!]
\begin{center}
\epsfig{file=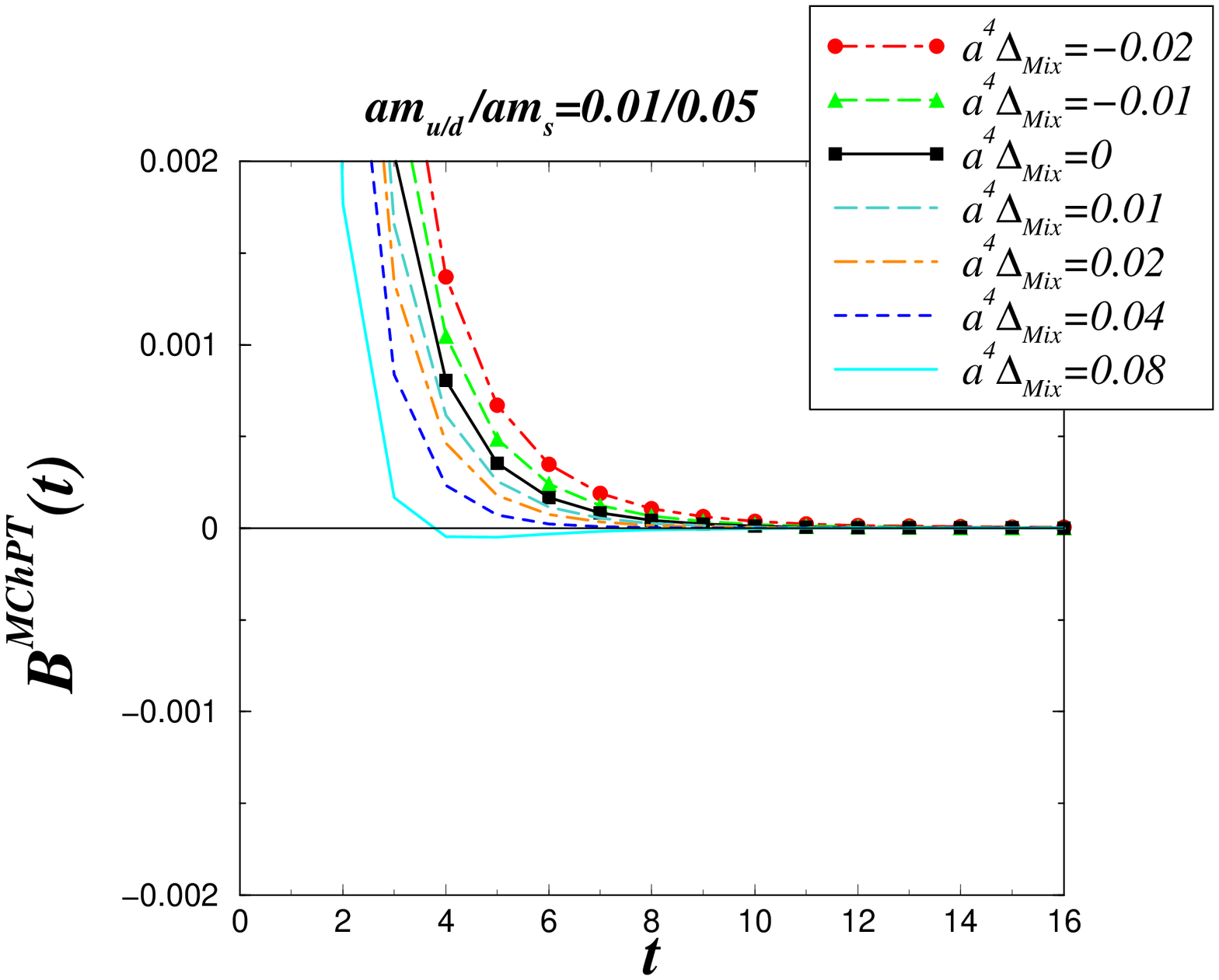,height=8cm}
\end{center}

\vspace{-0.8cm}

\caption{ \small The bubble contribution $B_{2+1}^{MChPT}(t)$ (\ref{Bmchpt}) for 
simulation with chiral fermions on $2+1$ MILC staggered coarse lattice 
with $am_{u/d}/am_s =0.01/0.05 $ and volume $20^3\times 32$. The valence and sea quark masses are tuned 
by matching $M_{val,val}=M_{U_I}$. }\label{fig.mchpt_tune2}
\end{figure} 

If the valence and sea quark masses are tuned by matching the masses of the valence-valence pion and the heaviest sea-sea pion ($M_{val,val}=M_{U_I}$), the bubble  contribution is relatively small and positive for 
$a^4\Delta_{Mix}\lesssim 0.04$ 
(see Fig. \ref{fig.mchpt_tune2}). 
Although 
the bubble contribution could be slightly negative for $a^4\Delta_{Mix}\gtrsim 0.04$, we expect the lattice scalar 
correlator to be positive in this case, since the small 
negativity in the bubble contribution is most likely out-weighted 
by the positive contribution $e^{-m_{a0}t}$ in (\ref{Ctot}).\footnote{The rough size of the $a_0$-exchange 
in the point-point 
scalar correlator can be estimated from $C_{a0}(t)=(16B_0^2 f_{a0}^2/m_{a0})
e^{-m_{a0}t}$ where $f_{a0}\simeq 0.08(3)$ GeV and 
$m_{a0}=1.58(34)$ GeV. }

Various choices of tuning the valence and sea quark masses have therefore 
large effects on the scalar correlator, although all the choices are 
equivalent 
in the continuum limit. UKQCD Collaboration in fact proposed to tune 
the valence and sea quark masses by considering the sign of the scalar 
correlator \cite{mixed_lat}, 
but the proposal has not been used in practice yet.    

\section{Conclusions}

The Nature of the observed scalar resonance $a_0(980)$ with $I=1$ is not 
revealed yet. The determination of $a_0$  mass on the lattice 
could indicate whether $a_0(980)$ is a conventional $\bar qq$ state or 
perhaps something more exotic. 

The $a_0$ mass is conventionally determined from the flavor 
 non-singlet scalar 
correlator on the lattice. However, 
 the interesting contribution $e^{-m_{a0}t}$    
to  the scalar correlator 
is accompanied by the contribution from multi-hadron intermediate states. 
The most important multi-hadron intermediate state 
is the bubble contribution $B$, which is the 
intermediate state with two pseudoscalars. In this paper we 
determined the size of the bubble contribution for simulations 
with three sea quarks of mass $m_u\!=\!m_d\!\not =\! m_s$. This offers a way 
to determine $m_{a0}$ by fitting 
the lattice scalar correlator to $C(t)=Ae^{-m_{a0}t}+B(t)$.

 We determined the bubble contribution for the case of simulations with  
 staggered sea and staggered valence quarks, 
as well as for simulations with staggered sea and chiral valence quarks. 
We found that the bubble contribution dominates the scalar correlator for 
small masses of $u/d$ quarks. In proper QCD, the 
lightest two-pseudoscalar 
intermediate state is $\pi\eta$ and the bubble contribution drops as   
$e^{-(M_\pi+M_\eta)t}$ at large $t$. The effects of taste breaking 
or mixed quark actions make possible also the intermediate state with mass 
$2M_\pi$, so the correlator drops as $e^{-2M_\pi t}$ at large $t$. 
This unphysical contribution to the scalar correlator vanishes 
only in the continuum limit. 
\begin{itemize}
\item 
We used Staggered ChPT for predicting the bubble contribution in case of 
staggered sea and staggered valence quarks. Its sign depends on the values 
of the input parameters, indicating the broken unitarity of the theory 
due to the $\root 4 \of{Det}$ trick. 
All the input parameters  have already 
been determined for the case of MILC configurations 
and they render positive bubble contribution.  
\item
The case of staggered sea and chiral valence 
quarks was explored using Mixed ChPT. The size and the sign of the bubble 
contribution depends on one unknown parameter $a^2\Delta_{Mix}$. 
The bubble contribution and the scalar correlator are  
negative (positive) for  most of the values 
$-0.02\leq a^4 \Delta_{Mix}\leq 0.08$    
if the valence and the sea quark masses are tuned by matching the masses of 
valence-valence pion and the lightest (heaviest) sea-sea pion.
The comparison of the point-point scalar correlator from lattice to 
our prediction offers a way to determine the unknown parameter 
 $a^2\Delta_{Mix}$. 
\end{itemize}
Our analytic prediction for the bubble contribution $B(t)$ 
will be helpful for the determination of $m_{a0}$ by fitting the 
lattice scalar correlator to $C(t)=Ae^{-m_{a0}t}+B(t)$. Such extraction of 
scalar meson mass is possible for moderate quark masses and times 
where  $e^{-m_{a0}t}$ is not negligible with respect 
to the bubble contribution.

\vspace{1cm}

{\bf  Note added:}

\vspace{0.2cm}

Some of the results given in the present paper were already presented 
in the conference proceedings \cite{sasa_lat05} by the same author.

\vspace{1cm}

{\bf  \large Acknowledgments}

\vspace{0.3cm}

We thank Kostas Orginos  
for pointing out the lattice simulations with 
chiral valence and staggered sea quarks, and for all the useful discussions. 
 It is a pleasure to thank 
Alan Irving, Carleton DeTar, Bob Sugar, Doug Toussaint and Claude Bernard 
for helpful discussions concerning the staggered simulations. We also kindly acknowledge Taku Izubuchi and Maarten Golterman for discussions on  
effective theories concerning   
mixed quark actions. This work is supported by the 
Ministry of Education, Science and Sport of the Republic of Slovenia.

\newpage

\appendix
\refstepcounter{section}
\section*{Appendix A: {\large Coupling of a scalar current to two pseudoscalars}}
\label{appA}
Here we derive the coupling of a point scalar current $\bar d(x)u(x)$ 
to a pair of pseudoscalar 
fields at the lowest order of ChPT. 
The effective scalar current can be determined from the dependence of the QCD Lagrangian and the Chiral Lagrangian on the spurion field ${\cal M}_{du}$, where ${\cal M}$ is quark mass matrix. The quark current $\bar d(x)u(x)$ is given by $-\partial {\cal L}_{QCD}/\partial{\cal M}_{du}$, so the effective current is represented by  $-\partial {\cal L}_{ChPT}/\partial {\cal M}_{du}$. The Chiral Lagrangian is 
\begin{equation}
\label{lagrangian}
{\cal L}_{ChPT}=\tfrac{1}{4}f^2{\rm Tr}[\partial^\mu U \partial_\mu U^\dagger]+\tfrac{1}{2}B_0 f^2{\rm Tr}[{\cal M}^\dagger U+U^\dagger {\cal M}]+...~,
\end{equation}
where $U=\exp[\sqrt{2}i\Phi/f]$ 
incorporates the $SU(3)_L\times SU(3)_R$ pseudoscalar field  matrix  $\Phi$, $f\sim 95$ MeV, while 
the dots indicate terms of higher order or terms which are independent 
of ${\cal M}_{du}$.  
The effective current is therefore equal to \cite{bardeen}
\begin{equation}
\label{appA_current}
\bar d(x)u(x)\sim -\frac{1}{2}B_0 f^2[U(x)+U^\dagger (x)]_{ud}=B_0 [\Phi(x)^2]_{ud}+ ...~,
\end{equation}
so the coupling of the current $\bar d(x)u(x)$ to a pair of pseudoscalar 
fields $\Phi$ is equal to the slope parameter $B_0=M_\pi^2/(2m_q)$. 
The result  (\ref{appA_current}) is derived within conventional ChPT here, 
but it applies also to various extensions 
of CHPT (Staggered, Mixed, Quenched and Partially Quenched) 
at the lowest order. In the extended case $\Phi$ in Eq. 
 (\ref{appA_current}) represents the field matrix of the
 extended   effective field theory. 

\appendix
\refstepcounter{section}
\section*{Appendix B: {\large Pseudoscalar propagators in Staggered ChPT}}

Here we list the propagators of pseudoscalar mesons in SChPT \cite{schpt} 
for three sea-quark flavors (with one or four tastes) 
in case of $m_u=m_d\not = m_s$. All the propagators in Euclidean space-time 
have the connected part 
\begin{equation}
\label{schpt_prop_con}
\langle \phi^b_{f f^\prime}|\phi^b_{f^\prime f}\rangle_{con}=\frac{1}{k^2+M_{ff^\prime_{~\!b}}^2}\quad,
\end{equation}
where $M_{ff^\prime_{~\!b}}$ is the mass of the un-mixed meson  
with flavor $\bar f f^\prime$ and taste $b=1,..,16$ 
\begin{equation} 
M_{ff^\prime_{~\!b}}^2=B_0(m_f+m_{f^\prime})+a^2\Delta_b~,\quad  M_{U_b}\equiv M_{\pi_b}\equiv M_{uu_{~\!\!b}}~,\ M_{S_b}\equiv M_{ss_{~\!\!b}}~, \ \Delta_5=0~.
\end{equation}
The taste-vector, taste-axial and taste-singlet mesons have also the 
disconnected part. 
The disconnected parts of $\langle \phi^b_{uu}|\phi^b_{uu}\rangle$, 
$\langle \phi^b_{dd}|\phi^b_{dd}\rangle$ and $\langle \phi^b_{uu}|\phi^b_{dd}\rangle$ \cite{schpt} are all equal to 
\begin{align}
\label{schpt_prop_dis}
\langle \phi^I_{uu}|\phi^I_{uu}\rangle_{disc}&=-\frac{4}{3}\frac{k^2+M_{S_I}^2}{(k^2+M_{U_I}^2)(k^2+\tfrac{1}{3}M_{U_I}^2+\tfrac{2}{3}M_{S_I}^2)}\quad {\rm for}\ m_0\to\infty\nonumber\\
\langle \phi^V_{uu}|\phi^V_{uu}\rangle_{disc}&=-a^2\delta^\prime_V\frac{k^2+M_{S_V}^2}{(k^2+M_{U_V}^2)(k^2+M_{\eta_V}^2)(k^2+M_{\eta^\prime_V}^2)}\nonumber\\
\langle \phi^A_{uu}|\phi^A_{uu}\rangle_{disc}&=-a^2\delta^\prime_A\frac{k^2+M_{S_A}^2}{(k^2+M_{U_A}^2)(k^2+M_{\eta_A}^2)(k^2+M_{\eta^\prime_A}^2)}~.
\end{align}
These hairpin couplings mix the flavor neutral mesons with $I,V,A$ tastes and the resulting mass eigenstates are \cite{schpt}
\begin{align}
M_{\pi_I}&=M_{U_I}~\ ,\ \ M_{\eta_I}^2=\tfrac{1}{3}M_{U_I}^2+\tfrac{2}{3}M_{S_I}^2~,\ \ M_{\eta^\prime_I}=\infty\qquad\quad\qquad {\rm for}\ m_0\to\infty \nonumber\\
M_{\pi_{V,A}}&=M_{U_{V,A}}~,\ M_{\eta_{V,A}}=m_{\eta}(C~a^2\delta_{V,A}^\prime)~\ , \ M_{\eta^\prime_{V,A}}=m_{\eta^\prime}(C~a^2\delta_{V,A}^\prime)
\end{align}
with $C=1$ in four-taste theory $2\cdot 4+4$ and $C=1/4$ in a single-taste theory $2+1$ \cite{schpt}. The functions $m_{\eta}$ and 
$m_{\eta^\prime}$  are defined as
\begin{align}
\label{meta}
m_\eta(\delta)&=\sqrt{\tfrac{1}{2}[M^2_{U_V}+M^2_{S_V}+3\delta-Z(\delta)]}~\\
m_{\eta^\prime}(\delta)&=\sqrt{\tfrac{1}{2}[M^2_{U_V}+M^2_{S_V}+3\delta+Z(\delta)]}\qquad {\rm with}\nonumber\\
Z(\delta)&=\sqrt{(M^2_{S_V}-M^2_{U_V})^2-2 \delta (M^2_{S_V}-M^2_{U_V})+9 \delta^2}\nonumber
\end{align}   
for the vector taste and analogously for the axial taste. 

\appendix
\refstepcounter{section}
\section*{Appendix C: {\large The sign of bubble contribution  in SChPT}}

Here we explore the sign of the bubble contribution $B^{SChPT}(t)$  
given in  
(\ref{Bschpt}). This is relevant since the positive definite 
bubble contribution to scalar correlator 
is expected in a proper unitary quantum field theory. 
We consider the unitary four-taste theory $2\cdot 4+4$
 and a single-taste theory $2+1$, where the later loses unitarity due to the 
fourth-root trick. 

Since the sign depends mostly on the values of the hairpin couplings 
$a^2\delta_{V}^\prime$ and $a^2\delta_{A}^\prime$, we concentrate on
 a special case 
 with $m_u=m_d=m_s$, $\Delta_b=0$ and 
$a^2\delta^\prime_V=a^2\delta^\prime_A\equiv a^2\delta^\prime$ for simplicity.
In this case 
\begin{align}
\label{proof}
\lim_{t\to \infty} B^{SChPT}(t)&=
\frac{B_0^2}{2\pi L^3}\biggl[\bigl(4C-\frac{4}{9}\bigr) 
\frac{3 \pi}{2 M_\pi^2}e^{-2M_\pi t}-8~a^2\delta^\prime~f_{dis}(C~a^2\delta^\prime)\biggr]\qquad \\
f_{dis}(\delta)&=\frac{\pi}{6~\delta~ M_\pi^2 M_{\eta^\prime}}\biggl[M_{\eta^\prime}e^{-2 M_\pi t}-M_{\pi}e^{- (M_\pi+M_{\eta^\prime}) t}\biggr]~,\quad M_{\eta^\prime}=\sqrt{M_\pi^2+3\delta}\nonumber
\end{align} 
with $C=1$ in four-taste  theory and $C=1/4$ in a single-taste theory.
It is easy to find a positive value of $a^2\delta^\prime$, which gives 
$B^{SChPT}_{2+1}(t)<0$ indicating the breakdown of unitarity 
due to the fourth root trick. On the other hand, we find that 
$B^{SChPT}_{2\cdot 4+4}(t)\geq 0$ for any values of 
$a^2\delta^\prime$, $M_\pi$ and $t$.  The positive definiteness of 
$B^{SChPT}_{2\cdot 4+4}(t)$ can be demonstrated by noting that 
$B^{SChPT}_{2\cdot 4+4}(0)>0$, $B^{SChPT}_{2\cdot 4+4}(\infty)=0$ and 
$dB^{SChPT}_{2\cdot 4+4}/dt<0$ for all $t$, 
which prohibits any extremum with a negative value. 
The positive definiteness of the scalar correlator in the four-taste theory 
is a consequence of the unitarity.

\end{document}